# SOXS NIR: Optomechanical integration and alignment, optical performance verification before full instrument assembly

M. Genoni*[a], M. Aliverti[a], G. Pariani[a], L. Oggioni[a], F. Vitali[b], F. D'Alessio[b], P. D'Avanzo[a], S. Campana[a], M. Munari[k], R. Zanmar Sanchez[k], A. Scaudo[a], M. Landoni[a], D. Young[o], S. Scuderi[k], P. Schipani[c], M. Riva[a], R. Claudi[d], K. Radhakrishnan[d], F. Battaini[d], A. Rubin[n], A. Baruffolo[d], G. Capasso[c], R. Cosentino[f], O. Hershko[e], H. Kuncarayakti[h,i], G. Pignata[l,m], S. Ben-Ami[e], A. Brucalassi[t], J. Achrén[p], J. A. Araiza-Duran[q,m], I. Arcavi[r], L. Asquini[a], R. Bruch[e], E. Cappellaro[d], M. Colapietro[c], M. Della Valle[c], M. De Pascale[d], R. Di Benedetto[k], S. D'Orsi[c], A. Gal-Yam[e], M. Hernandez Diaz[f], J. Kotilainen[h,i], G. Li Causi[g], L. Marty[c], S. Mattila[i], , M. Rappaport[e], D. Ricci[d], B. Salasnich[d], S. Smartt[o], M. Stritzinger[s], H. Ventura[f]

[a] INAF– Osservatorio Astronomico di Brera-Merate, via E. Bianchi 46, I-23807 Merate, Italy;
[b] INAF – Osservatorio Astronomico di Roma, Via Frascati 33, I-00078 M. Porzio Catone, Italy;
[c] INAF – Osservatorio Astronomico di Capodimonte, Sal. Moiariello 16, I-80131 Naples, Italy;
[d] INAF – Osservatorio Astronomico di Padova, Vicolo dell'Osservatorio 5, I-35122 Padua, Italy;
[e] Weizmann Institute of Science, Herzl St 234, Rehovot, 7610001, Israel;
[f] FGG-INAF, TNG, Rambla J.A. Fernández Pérez 7, E-38712 Breña Baja (TF), Spain;
[g] INAF– Istituto di Astrofisica e Planetologia Spaziali, via Fosso del Cavaliere 100, Roma, Italy;
[h] Finnish Centre for Astronomy with ESO (FINCA), FI-20014 University of Turku, Finland;
[i] Tuorla Observatory, Dept. of Physics and Astronomy, FI-20014 University of Turku, Finland;
[k] INAF – Osservatorio Astrofisico di Catania, Via S. Sofia 78 30, I-95123 Catania, Italy;
[l] Universidad Andres Bello, Avda. Republica 252, Santiago, Chile;
[m] Millennium Institute of Astrophysics (MAS), Santiago, Chile;
[n] ESO, Karl Schwarzschild Strasse 2, D-85748, Garching bei München, Germany;
[o] Astrophysics Research Centre, Queen's University Belfast, Belfast, BT7 1NN, UK;
[p] Incident Angle Oy, Capsiankatu 4 A 29, FI-20320 Turku, Finland;
[q] Centro de Investigaciones en Optica A. C., 37150 León, Mexico;
[r] Tel Aviv University, Department of Astrophysics, 69978 Tel Aviv, Israel;
[s] Aarhus University, Ny Munkegade 120, D-8000 Aarhus, Denmark;
[t] INAF– Osservatorio Astrofisico di Arcetri, Largo Enrico Fermi, 5, 50125 Firenze, Italy

## ABSTRACT

This paper presents the opto-mechanical integration and alignment, functional and optical performance verification of the NIR arm of Son Of X-Shooter (SOXS) instrument. SOXS will be a single object spectroscopic facility for the ESO-NTT 3.6-m telescope, made by two arms high efficiency spectrographs, able to cover the spectral range 350-2050 nm with a mean resolving power R≈4500. In particular the NIR arm is a cryogenic echelle cross-dispersed spectrograph spanning the 780-2050 nm range. We describe the integration and alignment method performed to assemble the different opto-mechanical elements and their installation on the NIR vacuum vessel, which mostly relies on mechanical characterization. The tests done to assess the image quality, linear dispersion and orders trace in laboratory conditions are summarized. The full optical performance verification, namely echellogram format, image quality and resulting spectral resolving power in the whole NIR arm (optical path and science detector) is detailed. Such verification is one of the most relevant prerequisites for the subsequent full instrument assembly and provisional acceptance in Europe milestone, foreseen in 2024.

* Contact: matteo.genoni@inaf.it

## 1. SOXS NIR SPECTROGRAPH

SOXS is a wide-band spectrograph for the NTT (it will be installed at one of the Nasmyth foci of the NTT) covering the spectral range from the UV to the NlR (350-2050 nm) in a single exposure. Its central structure (the backbone) supports two distinct spectrographs, one operating in the UV-VIS 350-850 nm and the other in the NIR 800-2050 nm wavelength ranges. Both spectrographs can operate at different (nominal design) resolutions according to the slit widths: R≈10000 with 0.5" slit, R≈4500 with 1" slit and R≈3300 with 1.5" slit. See for general project overview[1].

The near-infrared spectrograph, shown in Figure 1, is a cross-dispersed echelle, with R~5000 (for 1 arcsec slit), covering the wavelength range from 800 to 2050 nm with 15 orders (see for details[2]). It is based on the 4C concept, characterised by a very compact layout, reduced weight of optics and mechanics, and good stiffness. The spectrograph is composed of a double pass collimator and a refractive camera, an R-1 grating is the main disperser and a 3-prisms-based cross disperser. The detector is a Teledyne H2RG array operated at about 40K, 2048 x 2048 pixels, Pixel size 18 μm (see for details3). ESO NGC-I system is used for detector control and read-out.

The D-shaped Vacuum Vessel (VV), the thermal shield and the optical bench (which hosts all opto-mechanical assemblies and units) are made of 6082-T6 Aluminium. The VV includes one set of large KM used as an interface between the NIR spectrograph and the SOXS flange. The mass of the spectrograph is about 200 kg enclosed in a volume of about 900x700x400 mm (excluding the flanges and the cryo-vacuum elements connected to the flanges); the VV weights around 85 kg and its base is the only structural part of the system and it is 25 mm thick to minimize bending between ambient and operating pressure while all the walls are 20 mm thick. See for details[4,5].

The optical bench is supported by 3 flexures made of titanium alloy and G10; in order to ensure the stability of the slit location, a central pivot is mounted right below the Input-End assembly, which includes the stage used to select the different available slits.

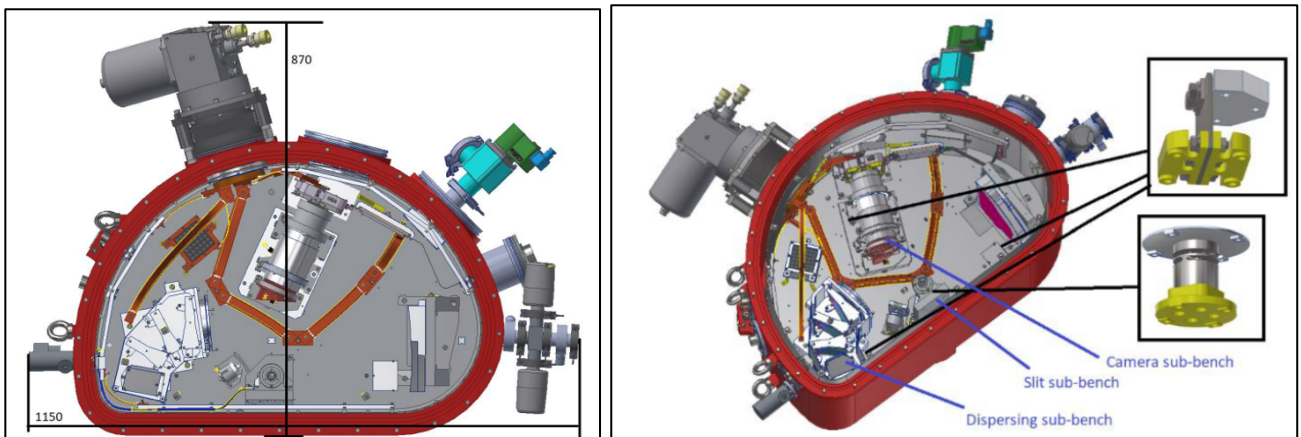

Figure 1. CAD overview of SOXS NIR spectrograph, including cryo-vacuum components. Right: detail of the 3 main optomechanical sub-benches and mechanical flexures units' location under the main bench.

This paper shows all the main assembly, integration, alignment, tests and verification activities performed in the INAF-Osservatorio Astronomico di Brera-Merate premises. All these activities have been successfully accomplished and the NIR spectrograph (along with all cryo-vacuum elements used in Merate premises) was packed and transported to INAF-Osservatorio Astronomico di Padova for final acceptance and integration with the whole SOXS instrument (see for details[6,7]).

## 2. WARM ASSEMBLY AND INTEGRATION ON TEST BENCH

The assembly and alignment of all the main optomechanical units mostly rely on mechanical characterization using Coordinates Measuring Machine (CMM, Coord3 Universal), these are:

- *The Input-End sub-bench*, which includes the slit, the pupil stop, the first lens and the first 45-deg mirror,
- *The Dispersing sub-bench*, which includes the collimator lens, the 3 prisms and the grating, (see Fig. 2),
- *The Camera sub-bench*, which includes the 3 camera lenses and the filter,
- *Main Collimator Mirror (CM) and Field Mirror (FM)* - both spherical mirrors.

The method foresees to measure reference points and surfaces of optomechanical elements with CMM and adjust (if needed) the element alignment until tolerances are met – comparing reference points and surface measurements with nominal values imported from CAD model. The sequence of optical element assembly and alignment (and features measurements with CMM) was defined case-by-case for the different units.

For example, for the Dispersing sub-bench, the 3 prisms starting from the one closer to the grating (P3 in Figure 3) have been mounted followed by the grating. In the case of prisms the assembly relies on the previously aligned reference blocks in front of each prism. Said alignment has been performed using dummy aluminum prisms. The block on the back has a spring to accommodate the thermal contraction during cooling down to operative temperature of about 150 K. After the full sub-bench assembly, the optical surfaces were verified with CMM as shown in Figure 3.

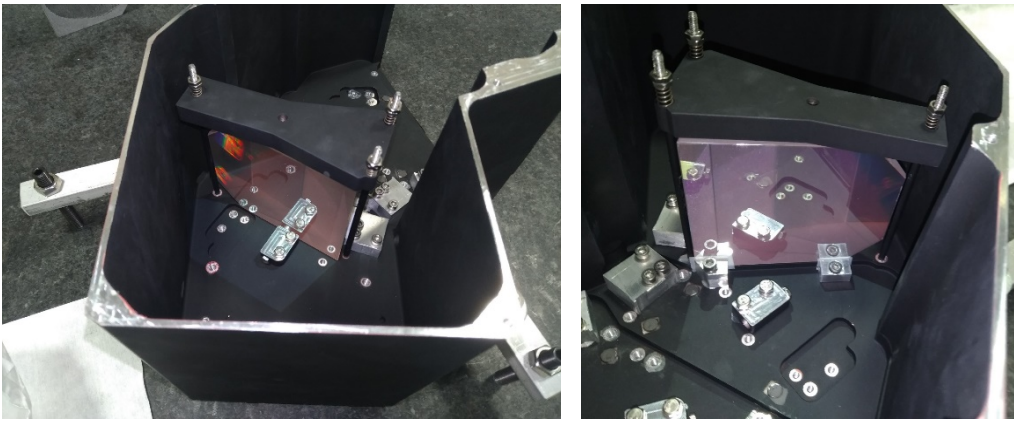

Figure 2. Assembly of the prisms into the Dispersing sub-bench. Detail of the reference blocks (with PTFE protections) and spring block for thermal contraction experienced by the prisms during the cool-down to operative temperature (about 150K).

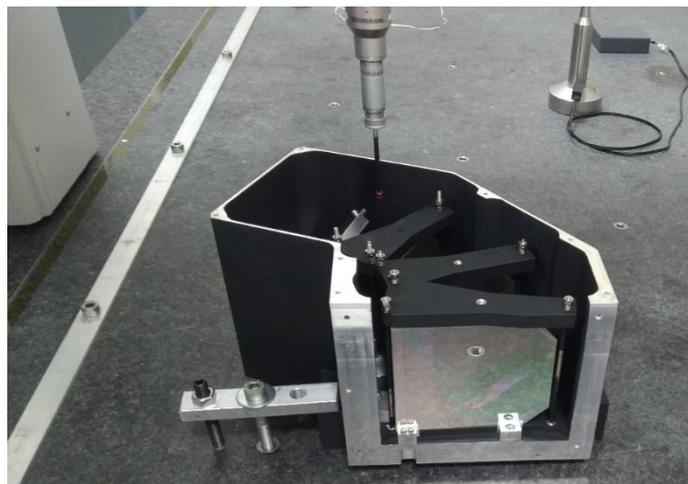

Figure 3. Alignment verification of the 3 prisms and grating into the Dispersing sub-bench, via CMM.

Then, all the NIR optomechanical sub-benches and units were integrated onto the *"Warm Test Bench"* to perform optical tests (IQ, orders identification and curvature) before integration into Vacuum Vessel. The *"Warm Test Bench"* has the same reference pins, which were used as reference for installation, as the NIR Optical Bench inside the VV. As shown in Figure 4, on the left the *Collimating Mirror* with its mask, on center-bottom the *Camera sub-bench* (3 camera lenses with no baffles in this picture), on center-top the *Input-End sub-bench* and the spherical *Field Mirror,* on the right of the *Dispersing sub-bench*. In the focal plane a CMOS 2.2 μm pixels is used instead of the science grade NIR detector.

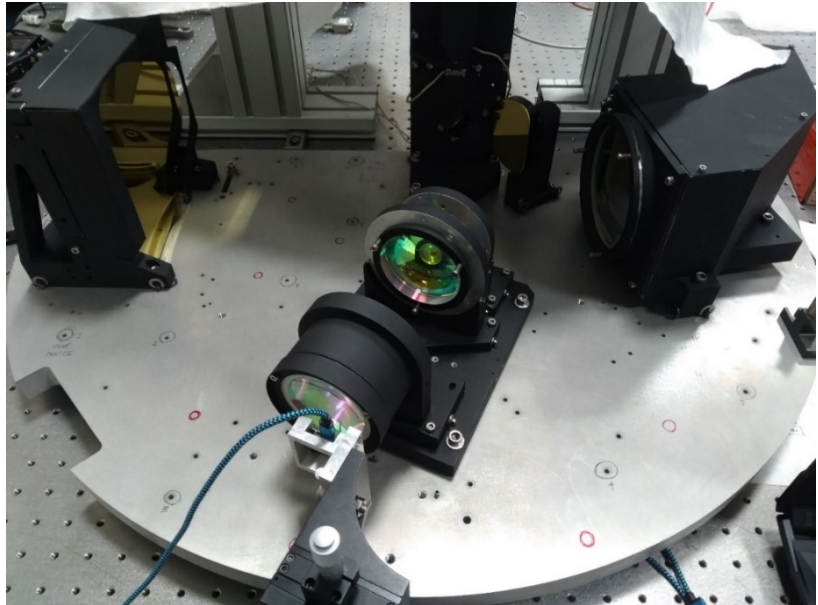

Figure 4. All NIR optomechanical sub-benches and units mounted onto the "Warm Test Bench".

Being the spectrograph pre-aligned warm condition, a CMOS 2.2 μm pixels (see Figure 4) was used to record at least images of different portions of the diffraction orders 21, 22, 23, and 24. Redder orders cannot be detected because of the CMOS cut-off. The performed tests were: i) diffraction orders and wavelength identification, ii) Inter order separation and order curvature, iii) image quality and resolution elements FWHM.

An example of the image of an equivalent 0.5" seeing in passing through the 0.5" slit aperture, for wavelength 849.536 nm, is shown in Figure 5. A residual astigmatism can be noticed, resulting in a non-perfectly circular shape; this is caused by the WFE residual introduced by the reflection grating; despite this residual astigmatism, the FWHM is as expected.

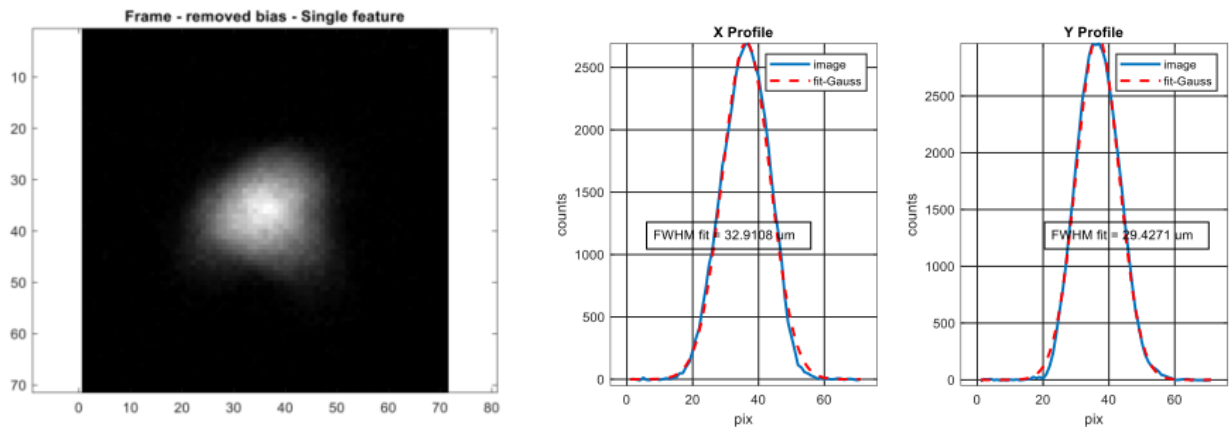

Figure 5. Image of the 849.536 nm feature of the Ne calibration Lamp. FWHM and Gaussian fit of the main (left) & cross dispersion (right) binned profiles confirm compliance with theoretical values.

## 3. OPTOMECHANICS INTEGRATION IN THE VACUUM VESSEL

After all the tests were done in warm conditions on the *"Warm Test Bench"*, all optomechanical units were integrated inside the NIR Vacuum Vessel using nominal spacer w.r.t. mechanical reference pins, to pre-align all degrees of freedom. The alignment of sub-benches and units is re-checked (ad adjusted when necessary) using reference points measured with an articulated arm (FARO-Arm EDGE) since inside the VV many points are not accessible with the CMM. The measures are imported into the CAD model and alignment residuals are checked to be in specification.

As an example, Figure 6 shows the measurements of three planes (red planes overlaid on the mechanical structure) of the Input-End sub-bench: front, lateral and top. From these measures all 6 d.o.f. are verified – in Figure 6 specific case of the top plane residual along vertical direction is shown (being < 100um tolerance); the same was done for all d.o.f. of all sub-benches and units.

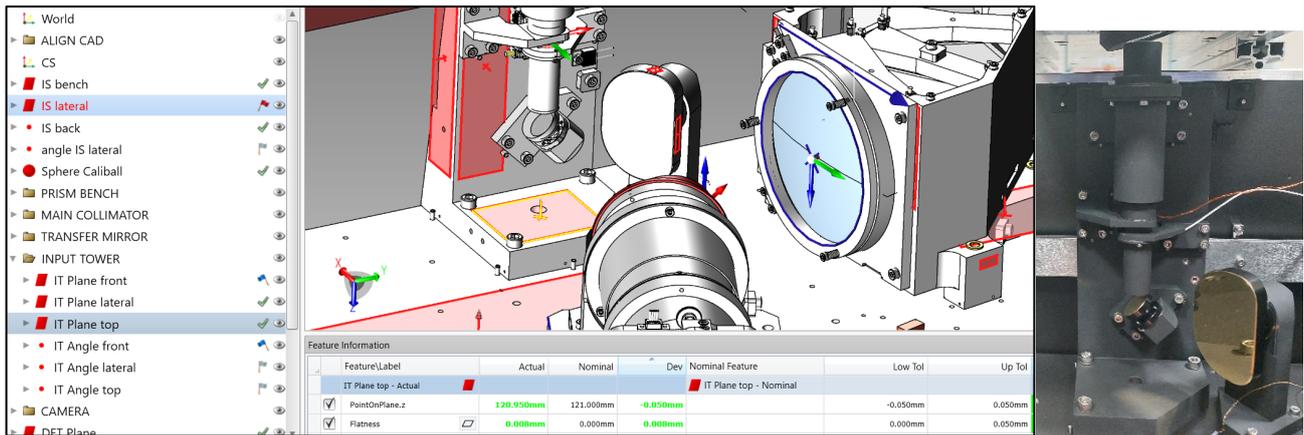

Figure 6. Mechanical measurements of the Input-End sub-bench imported in the CAD model. Residuals are in specifications. On the right, is a picture of the aligned Input-End sub-bench.

The optomechanical assembly used to replicate the beam projected inside the NIR spectrograph from the SOXS Common-Path subsystem is the Common-Path Simulator and in the final use-case, it was mounted on the VV top cover (see Figure 10 in sec. 4). In the tests done before installing the SG Detector, the VV top cover was not used to allow item accessibility and thus the CP-Simulator was mounted on dedicated structure on VV. From the top view of Figure 7, it is possible to see the structure and the CP-simulator bench holding the two off-axis parabolas with the diaphragm setting the pupil size. The maximum position residual, w.r.t. theoretical value, is about 70 μm (focus position); while the angle is about 1 arcmin.

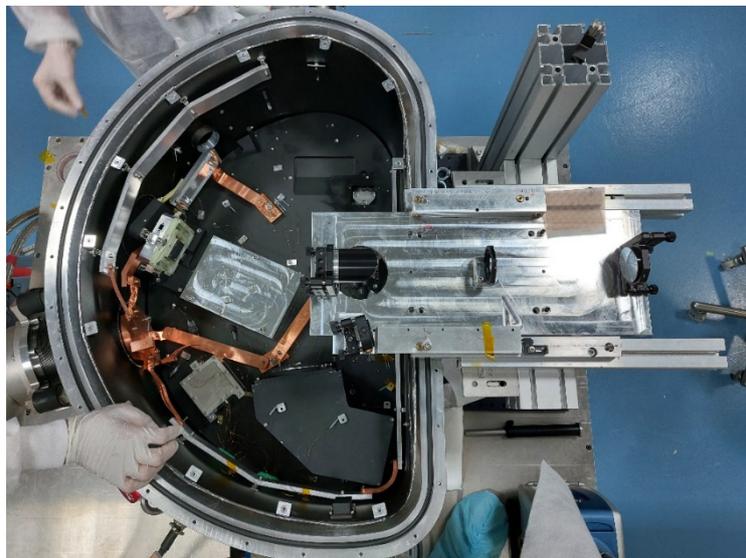

Figure 7. Integration of All units into the VV. The Collimator Mirror and Camera barrel are not already installed. See the CP simulator (composed of 2 off-axis parabolas with a diaphragm to set the pupil size) and its bench on the right.

All the optical tests mentioned in section 2 (i.e. image quality, orders identification and curvature) were re-done before installing the NIR SG Detector, again with the same CMOS 2.2 μm pixels – see on the right of Figure 8, at the focal plane of the camera (all the Camera baffles between the lenses are now installed). After successful verification (as described in section 2), the SG Detector was installed onto its mechanical mount and then integrated onto the optical bench with pre-amplifier and the whole read-out circuit, see the bias and clock PCBs on the right of Figure 9.

The detector frame (light grey mechanical frame) is aligned w.r.t. the mechanical mount using three regulation screws with a v-groove pad for high repeatability (ESO design). Additional baffles around the last lens of the camera, G-10 connections between the pre-amplifier frame and the detector mechanical mount as well as the thermal connection between the Cold-Head and the detector frame are shown in Figure 9.

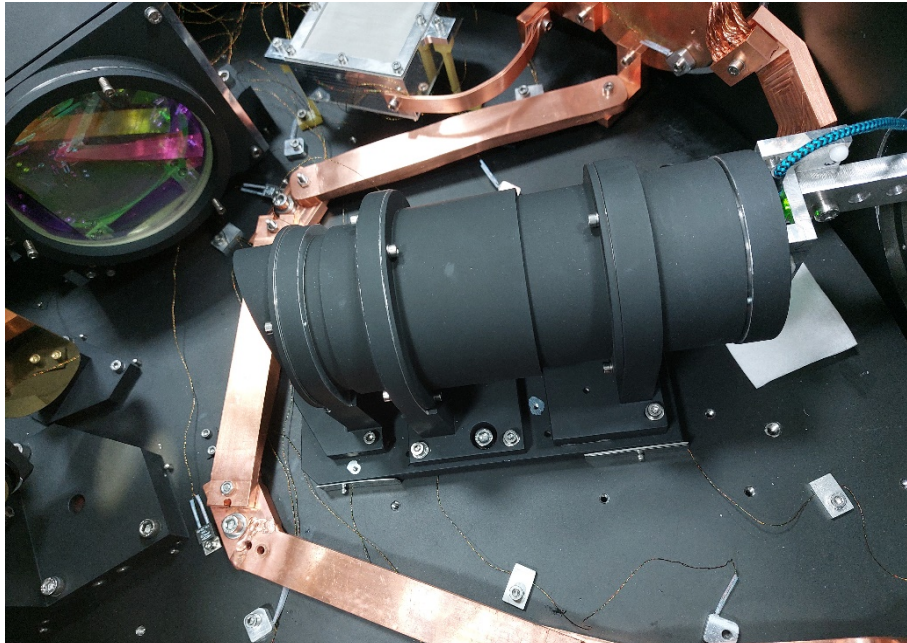

Figure 8. Camera barrel detail with CMOS sensor at the focal plane for warm optical tests repeated with all units in the VV.

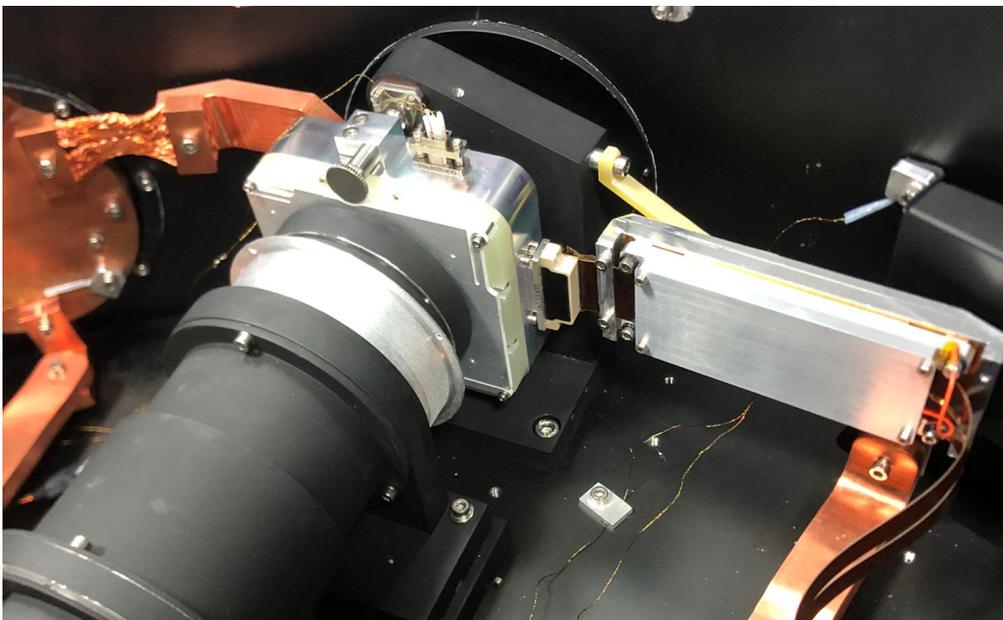

Figure 9. After warm optical tests in the VV, the final NIR SG Detector Installation with additional Baffling onto the camera Barrel.

## 4. SET-UP FOR OPTICAL TESTS AND PERFORMANCE VERIFICATION

Two set-ups have been used for the spectrograph optical tests. They differ in the source position on the CP simulator, which allows for different entrance slit illumination. The one shown in Figure 10 homogeneously illuminates the entrance slit, emulating a flat spatial source onto the CP pupil. The second one, see Figure 11 and Figure 12, projects a fiber image onto the entrance slit plane with a magnification factor of 0.5. Both of them project a beam at approximately F/6.7.

The first set-up is the one used to evaluate the spectrograph image quality and resolving power since it emulates the operational observing condition which will be slit-size limited.

The calibration pen-ray lamp (in the specific case of all the tests here performed it was an Xe lamp since it provides a good number of spectral features distributed for almost all SOXS NIR diffraction orders) is positioned onto the CP simulator bench before the CP pupil diaphragm, a diffusing glass is placed almost onto the pupil diaphragm. The 90-degree off-axis parabola (on the left side in Figure 10) projects the beam towards the spectrograph entrance window.

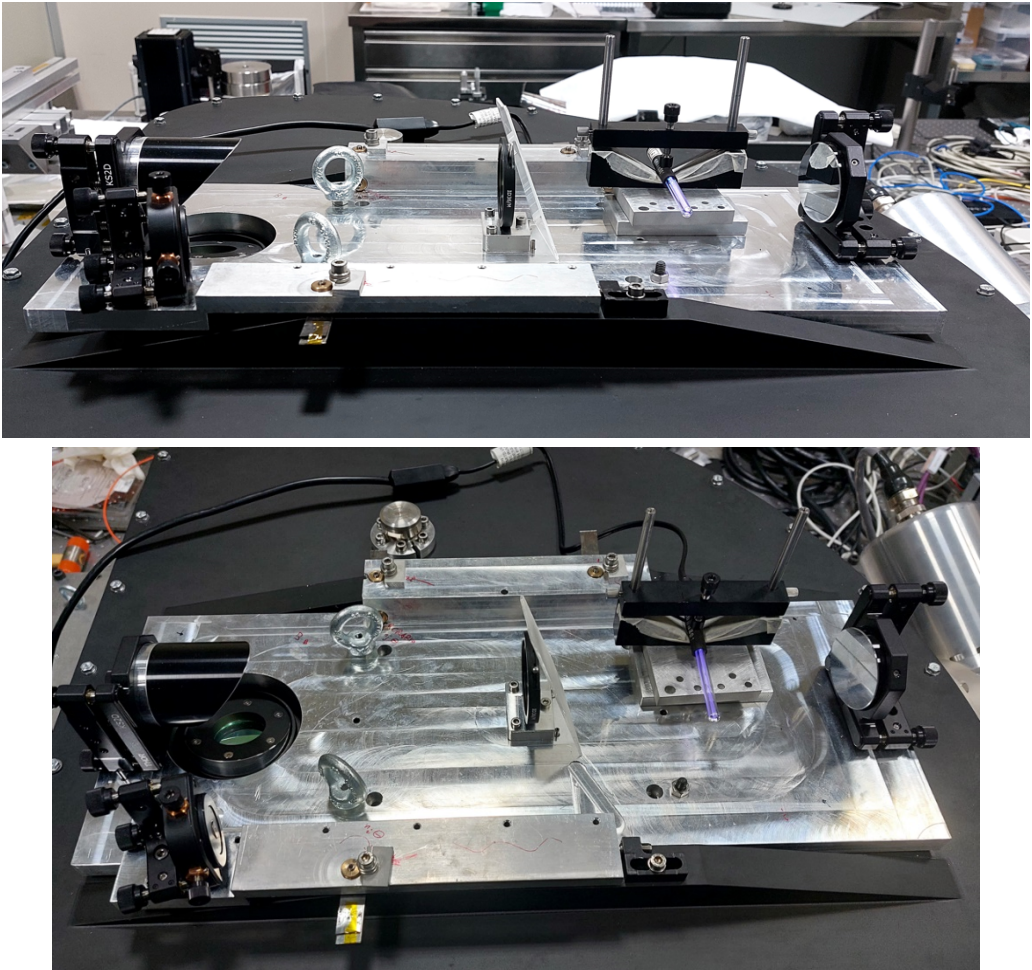

Figure 10. Example of the CP simulator pupil illumination set-up. Side and top views.

The second one has been used to evaluate the through-focus of the NIR spectrograph by moving the position of the input fiber (see Figure 12), in order to retrieve the conjugated and equivalent detector motion to place it onto the spectrograph focal plane.

In this setup, the pen-ray Xe lamp is positioned inside the integrating sphere and an optical fiber transfers the light to the input of the CP simulator; all other ports of the integrating sphere are properly closed.

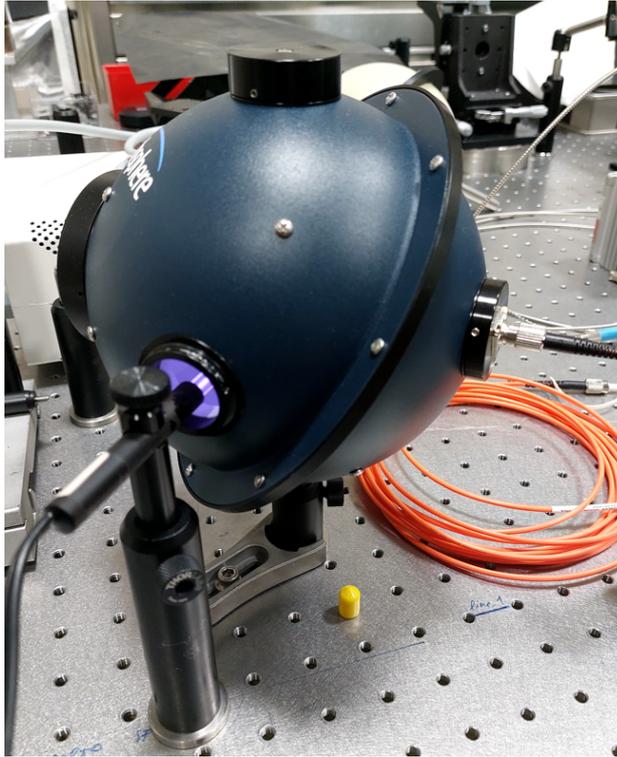

Figure 11. Pen-ray Xe lamp mounted on an optical post, Integrating Sphere and optical fiber set-up. The fiber transfers the light to the input of the CP simulator (see Figure 12).

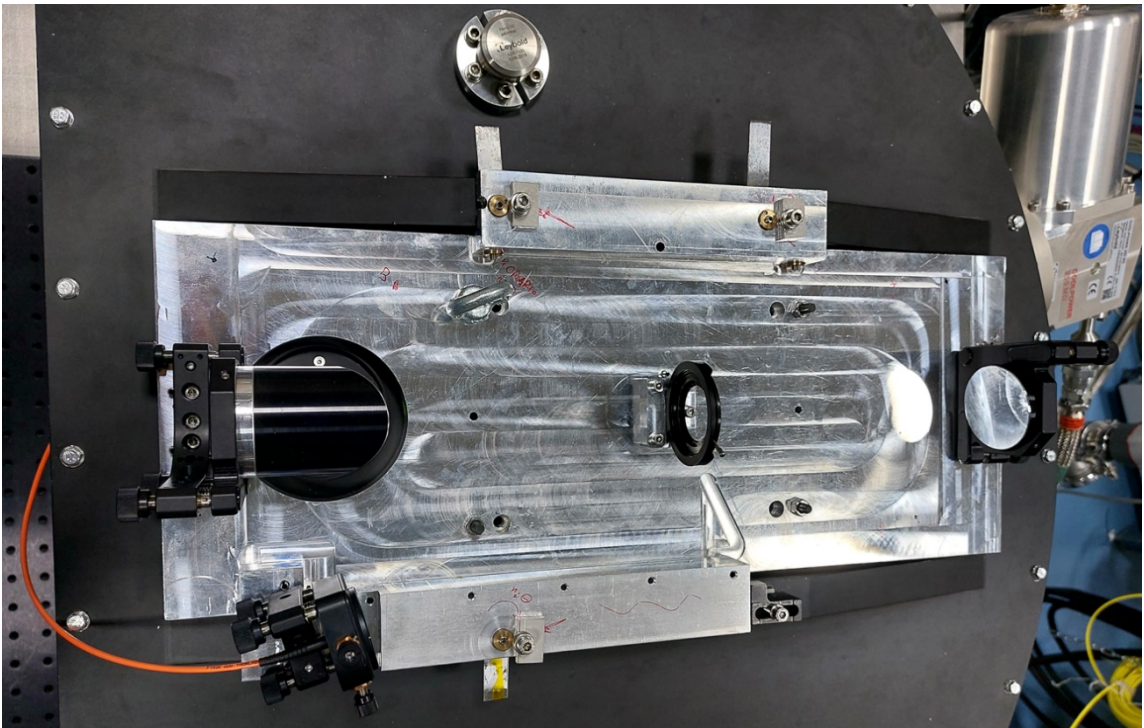

Figure 12. Top view of the CP simulator mounted onto the closure of the VV with optical fiber at the input (bottom left).

## 5. PERFORMANCE VERIFICATION

### 5.1 Functional Verification

The only moving function of the NIR spectrograph is the stage which exchanges and selects the NIR entrance aperture (0.5'', 1.0'', 1.5'', 5''. Single pinhole 0.5'' and multi pinhole). The functional verification was done by checking the spectral features position while moving the stage (for all different slits – see Figure 13). The stage has been moved exactly of the theoretical average distance between the center of the two slits (0.5 and 1.0 arcsec), which is 1.9 mm. The features are in the same position on the detector, to better than about 0.5 pixels (≈ 10 μm onto the entrance slit plane); this means that the motion given by the stage is correct and the residual is just the not perfect average distance between different slits.

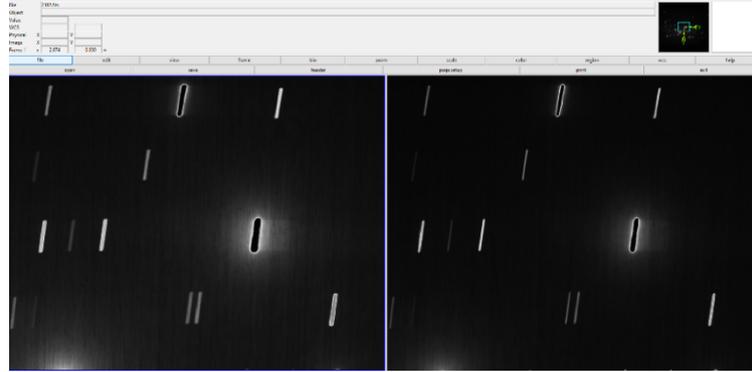

Figure 13. Comparison of some spectral features position with 0.5 arcsec (on the right) and 1.0 arcsec (on the left) slits.

### 5.2 Spectral format, Image quality and Resolving power

The lamp used is Xenon, since it provides a good number of spectral features distributed for almost all NIR diffraction orders; 25 wavelengths in the FSR have been identified and used for the analysis, specifically 5 orders and 5 wavelengths per order, as shown in Figure 14.

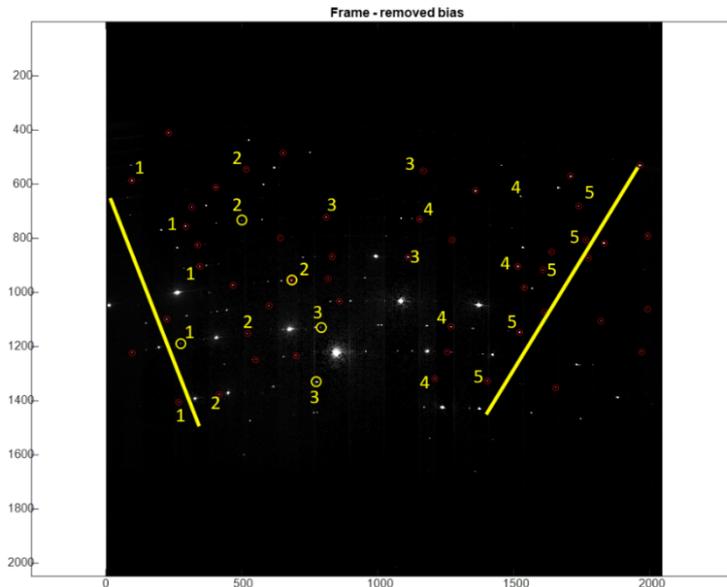

Figure 14. Example of single pin-hole frame with an indication of 25 wavelengths used for analysis (yellow numbers). Red circles are other features identified in the spectral format.

The identified wavelengths are in agreement with the expected positions along diffraction orders, emulated with the SOXS instrument simulator[9], given the final predicted optical elements alignment in cryo-vacuum conditions; this also indicates that the different diffraction orders wavelength coverage and overlap is in agreement with nominal design conditions (the full verification using the dedicated pipeline recipes is ongoing for the upcoming PAE process scheduled between July and October 2024).

For all the 25 wavelengths indicated in Figure 14, the FWHM in main dispersion was retrieved from the fitted 1D profile. Specifically, for the long slits, the FWHM was evaluated per each 1D line/row of the image and the mean values is considered for all the features (checking that the variation of the FWHM along the slit is negligible). Concerning the Single Pin-Hole, the image was binned along cross dispersion direction and then the FWHM of the 1D profile was evaluated (FWHM-X); binning along main dispersion was done for computing the FWHM in cross dispersion direction (FWHM-Y). In addition for all the 25 spectral features, the maximum counts and total integrated counts in the image were computed for different detector integration time (DIT) values; the variation of counts along with the DIT values, was exploited to identify possible saturated features (for the specific DIT) which might not be reliable in the analysis of the image quality. Figure 15 shows the FWHM-X and Peak intensity for the 1.0 arcsec slit, combining different exposures to analyse the 25 features when the peak intensity is about 30000 counts (saturation is above 50000 counts), which is usually a reliable indication of performance.

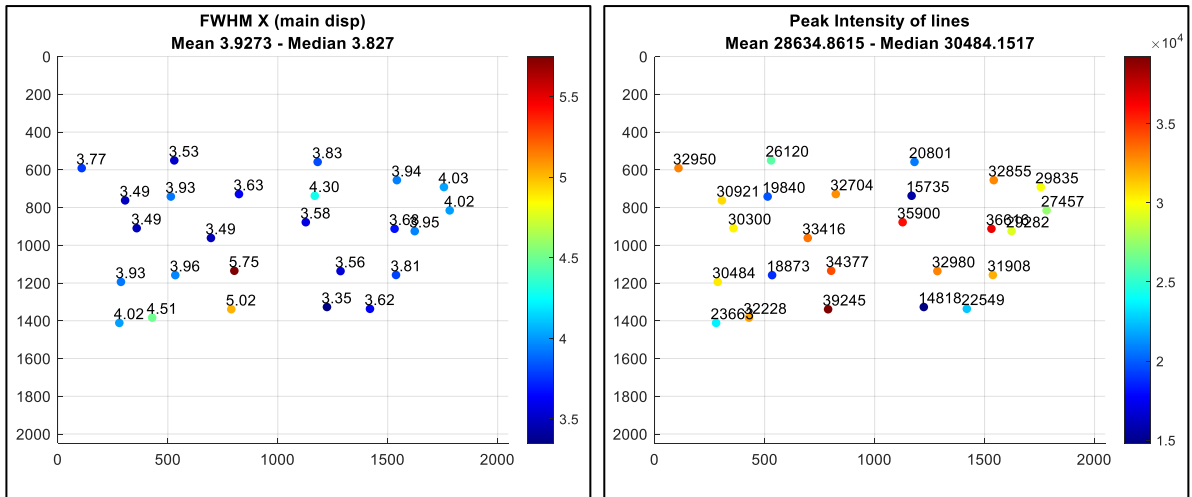

Figure 15. Slit 1.0 arcsec, combining different DITs avoiding lines saturation. Left plot: FWHM-X, right plot: counts.

The Resolving Power, R, is computed using the formula: $R = \lambda/\delta\lambda$. Where $\delta\lambda$ is derived from the FWHM-X value computed for the 25 wavelengths taking into account their linear dispersion. The computation presented in Figure 16, uses the FWHM-X from the non-saturated lines at about 30,000 counts shown in Figure 15. The relevant TLR asks for R>= 3500 for the whole spectral coverage of the NIR spectrograph. From the current analysis it is possible to conclude that R measured is compliant with the TLR and in agreement with theoretical design values, see Figure 16.

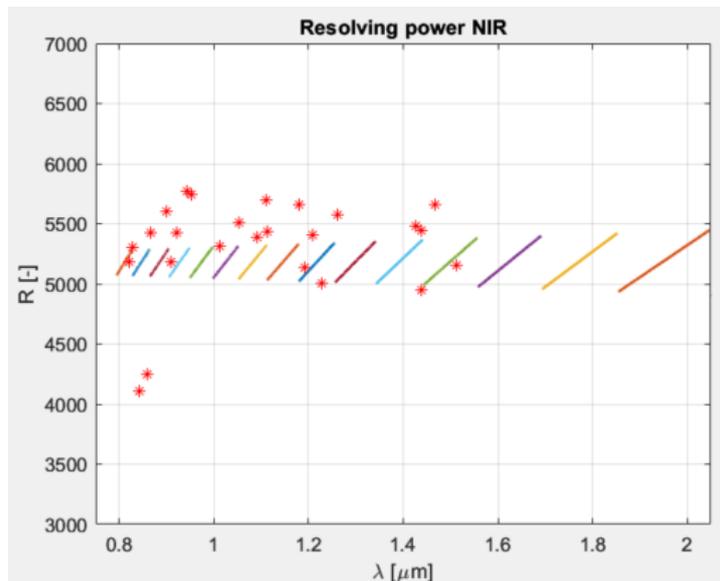

Figure 16. Resolving Power plots of the 1 arcsec slit, related to FWHM in main dispersion computed in Figure 15.

The obtained PSF FWHM in both main dispersion and cross dispersion direction were compared with theoretical values exploiting the SOXS instrument simulator as described in a parallel proceeding[9]. The results shown there, demonstrate a very good match between real PSF FWHM values with theoretical ones in most of the echellogram, while a performance degradation (especially in cross dispersion direction, due to known spectrograph astigmatism) can be seen in the bluest orders and at the edges of the FSR (coherently with results reported in Figure 15).

**5.3 Flexures and spectral format stability**

Flexure and spectral format position stability test was done comparing the Single Pin-Hole frames taken with the VV in horizontal position, and frames taken with VV in vertical position. This is done by tilting the tilting table (see Figure 17).
The flexure verification was done by computing the centroid X and Y of the 25 wavelengths used for image quality and resolving power analyses, for both horizontal and vertical frames and comparing the values (i.e. centroid from vertical frames w.r.t. centroid from horizontal frames) of each spectral features.
Results are presented in bubble plots of Figure 18. The computed feature motion is given, in pixel unit, as number in the middle of each bubble. The mean value is below 1 pixel in both directions and the maximum absolute motion is 1.46 pixels (computed as RSS of X and Y shifts per each spectral feature).
It can be noticed a trend along dispersion direction in each analysed order, with larger absolute shifts in the red part of each order (i.e. at larger wavelengths of each order). The shifts along the main dispersion direction would not induce a relevant resolving power decrease and the maximum shift along the cross dispersion direction (about 1.45 pixels) is about 3.3% of the slit length projected onto the detector, thus not producing critical effects for frame calibration.
Nevertheless, it shall be underlined that such tested flexure due to VV orientation variation is an extreme condition that will not happen during common observations of 20 minutes to 1 hour integration time at the telescope and the spectrograph rotation will be as well around a different axis being mounted on the instrument flange.
Additional and fully representative flexure tests are planned with the NIR spectrograph mounted and aligned to the SOXS instrument for the system verification before PAE in Padova premises.

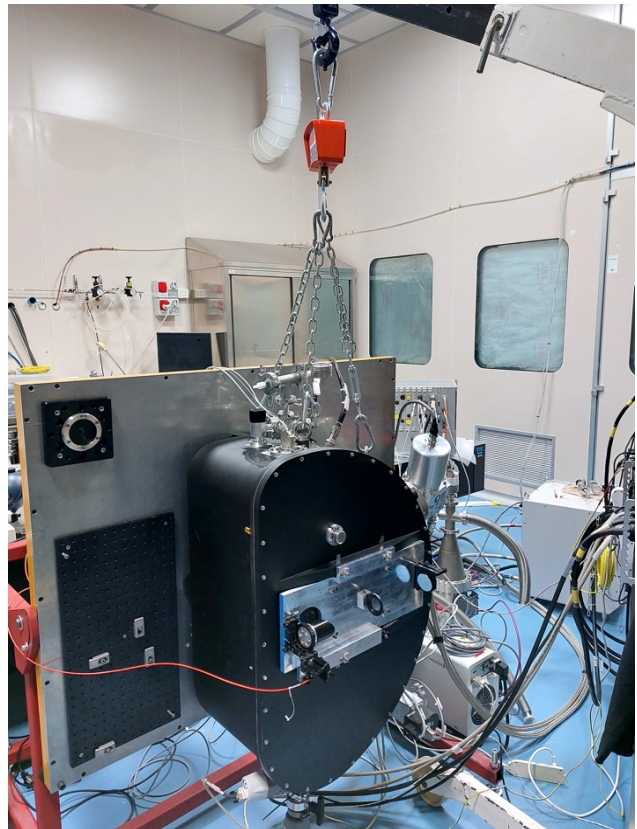

Figure 17. NIR VV vertically oriented to perform flexure tests.

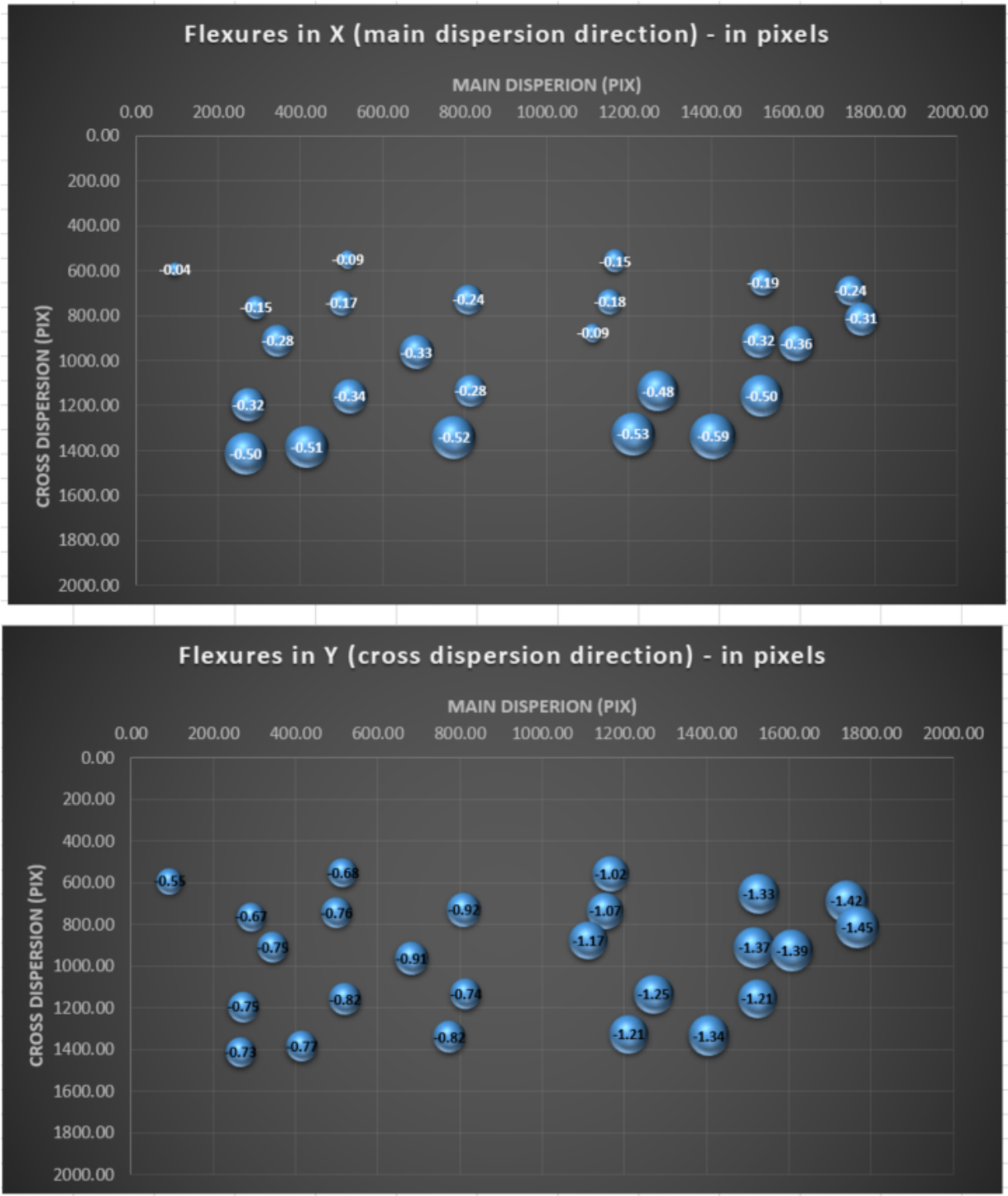

Figure 18. NIR spectral format stability verification – horizontal –vs- vertical condition.

**5.4 Thermal background noise and Detector linearity**

Despite the light tight design of the thermal shield inside the VV, initial dark tests spotted high counts level and distribution onto the detector compatible with residual unwanted thermal noise. The camera is equipped with a thermal filter cutting above ~2.1 μm; therefore the residual noise was for the wave-band below such cut-off. A long series of tests were done to understand thermal background sources in order minimize the noise and to keep it below 0.05 e-/pix/s as required. These tests were compared with optical ray-tracing simulations done with Optics Studio in Non-Sequential mode in order to spot and block all possible paths of thermal radiation reaching the detector. A very useful test that finally allowed us to identify all light leaks from the thermal shield was to introduce inside the VV a bright light source, to close the top thermal shield and look for light coming out from the space between the thermal shield and the VV. The final additional installed baffles are:

- around the cold-head to block radiation entering from the VV lateral structure;
- between the sorption pump and Dispersing-Camera sub-benches, to avoid residual thermal radiation reaching the Field Mirror corner/chamfer and reflected back towards the camera;
- aluminium foil to cover the residual interspace between the top and the side thermal shield; the foil is folded once the top shield is closed;
- aluminium foil around the camera L3 mount and detector frame to avoid possible reflection from the frame and back reflection from L3 reaching the detector.

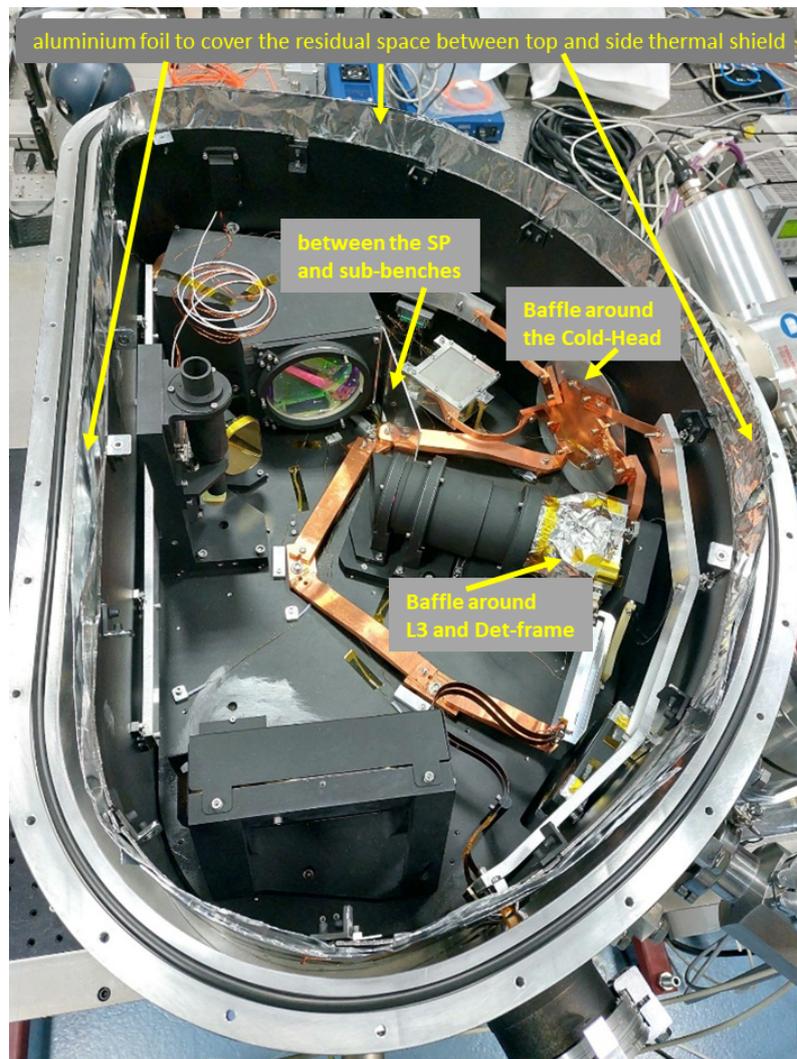

Figure 19. NIR Spectrograph additional installed baffles to minimize thermal noise. See figure texts for locations and description.

After all baffling installation, a series of dark frames was obtained at different temperatures of the vacuum vessel (VV), spanning from 156 to 142 K, with the detector being always at T ~ 42 K. The detector read-out procedure was the Up-the-

Ramp mode and the derived gain 2.3 e-/DN (digital number); the DIT was 1200 seconds for all frames. Results are plotted in Figure 20, where it can be derived that the requirement is met for VV temperature below 148/147 K with a possibly final operating temperature (which will be finally set during system tests in Padova) at 145 K.

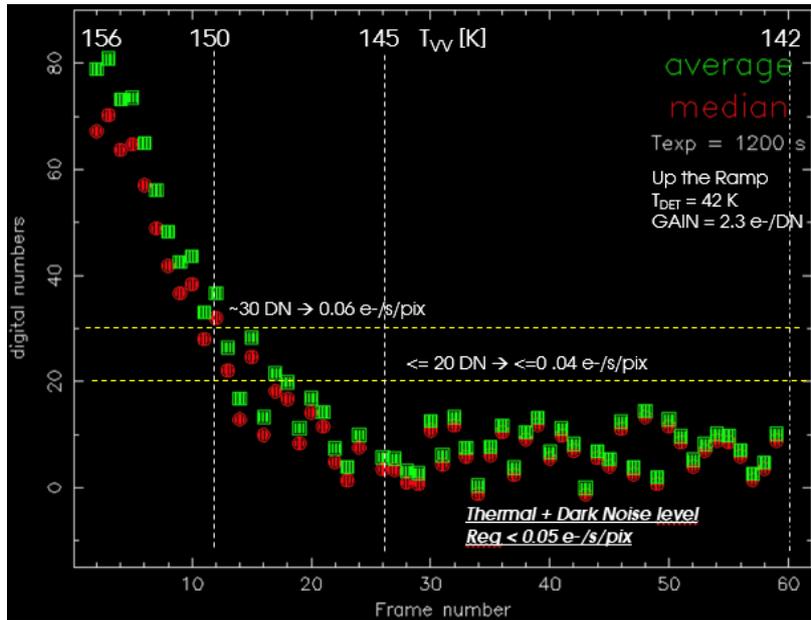

Figure 20. NIR thermal and dark noise level plot, for VV temperature range from 156 to 142 K and detector at about 42 K. Both average (green squares) and median (red circles) measures are shown. The requirement is met for the region below 148/147 K.

A linearity test has been carried out by taking the spectra of a halogen (QTH) lamp with the 5" slit and increasing exposure times. The result is that linearity start to get lost around 3e4 counts, while saturation occurs at around 5e4 counts. No significant difference is observed by carrying out the test with the different detector reading modes (Correlated Double Sampling and Up The Ramp).

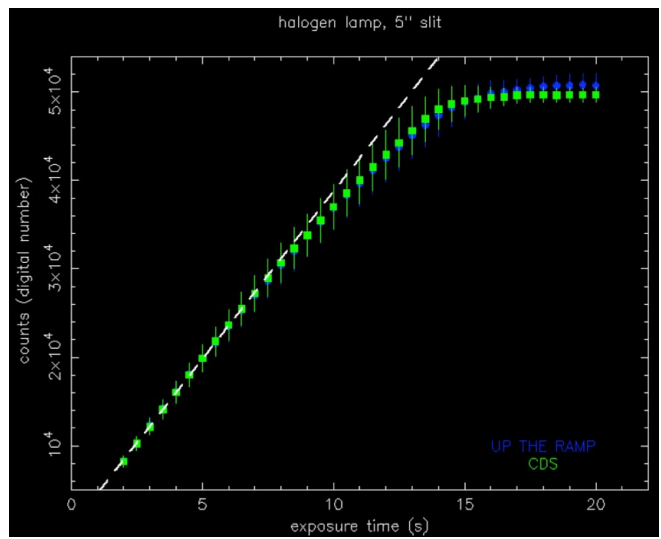

Figure 21. NIR detector linearity test taking spectra of a halogen (QTH) lamp with the 5" slit linearity start to get lost around 3e4 counts, while saturation occurs at around 5e4 counts.

## 6. CONCLUSIONS

This paper presented all the main assembly, integration, alignment, tests and verification activities performed in the INAF-Osservatorio Astronomico di Brera-Merate premises. The sequence of optomechanical sub-benches and units' assembly and alignment, as well as their installation onto a *"Warm Test Bench"* to perform tests before integration into the VV was summarized. The approach used to integrate and for the alignment check of all sub-benches into the optical bench inside the VV with articulated arm was described, giving the example of performance achieved with the Input-End sub-bench (performance representative for all the sub-benches and units).

The functional and performance verification tests show that the spectral format, image quality and Resolving power, and thermal+dark current noise level are in agreement with design and compliant with requirements. For what concerns spectral format and image quality comparison between real data and simulated frames, generated with the SOXS End-to-End simulator (see details here[9]), was exploited to further confirm the goodness of performance results.

Flexure and spectral stability tests were done with the VV in horizontal and vertical position, from which good results have been obtained.

The NIR spectrograph with all CryoVacuum items was packed and shipped to INAF-Padova premises for the full SOXS instrument assembly and system AIT/V (see references[6,7,8]), where the final full functional, performance and flexure tests are planned.